# Observation of Iron Oxide-to-Nitride Conversion via Liquid–Liquid Phase Separation in High-pressure Borate Melt

Yu Tao[1], Depu Liu[1], Chunyin Zhou[2], Xu Jia[1], Jingyi Liu[1], Xue Chang[1], Yangbin Wang[1], Yipeng Wang[2], Duanwei He[1], Li Lei[1*]

1. Institute of Atomic and Molecular Physics, Sichuan University, Chengdu 610065, China

2. Shanghai Advanced Research Institute, Chinese Academy of Sciences, Shanghai 201204, China

**Abstract:** High-pressure chemistry provides a powerful route to materials that are inaccessible or difficult to synthesize under ambient conditions. However, high-pressure chemical reaction processes and mechanisms remain largely unexplored because of the challenges associated with *in-situ* characterization under high pressure and high temperature, particularly within the deeply enclosed sample environment of a large-volume press. Here, we employ the state-of-the-art real-time synchrotron X-ray radiography to image a high-pressure chemical reaction at 5.4 GPa and ~1700 K within a large-volume press. Using $Fe_2O_3$ and BN as precursors, we capture the complete dynamic metal oxide-to-nitride conversion and show that it differs fundamentally from conventional solid-state diffusion-controlled nitridation. Synchrotron X-ray radiography clearly revealed that this conversion involves a two-stage liquid-liquid separation process (LLPS), including fluid nitrogen and fluid Fe-N alloy. On the basis of these observations, we propose a nitrogen-driven mechanism for LLPS in borate melts. Specifically, nitrogen reduces metal cations in the borate network, altering their coordination environments and triggering a substantial reorganization of the melt structure. This coordination-induced restructuring destabilizes the borate melt and promotes the LLPS of fluid Fe-N alloy. Our *in-situ* observations suggest a general pathway for high-pressure metal oxide-to-nitride conversion. This study provides a direct visualization of a pressure-enabled chemical reaction that is inaccessible under ambient conditions, offering fundamental insight into how high pressure reshapes chemical reaction pathways and enables the synthesis of metal nitrides.

High pressure significantly affects the interatomic interactions, thereby changing the paradigm of chemical reactions. In the last decades, novel materials unattainable or difficult under ambient conditions, including superhard materials[1,2], high-temperature superconductors[3,4], high-energy-density materials[5,6] and transition metal nitrides[7,8], were synthesized by high-pressure chemistry methods. High-pressure chemistry has gradually emerged as a new branch within the field of chemical research[9,10]. However, research in high-pressure chemistry is mainly focused on the intriguing properties of reaction products, while studies on high-pressure chemical reaction processes and mechanisms remain largely unexplored. This limitation arises primarily because most high-pressure reactions occur under high temperature, where direct observation of chemical processes inside high-pressure apparatuses remains highly challenging. The challenge is particularly acute in large-volume presses, where chemical reaction occurs within a highly confined and sealed reaction volume under high pressure and high temperature. To the best of our knowledge, no previous study has directly visualized, through in-situ characterization, the dynamic evolution of a chemical reaction occurring under high pressure.

In this study, we investigated the mechanism by which metal nitrides form under high pressure. Metal nitrides represent one of the most extensively studied classes of materials in high-pressure chemistry. High pressure effectively overcomes the chemical inertness of nitrogen and promotes reactions between nitrogen and metals, leading to a variety of nitrides with attractive mechanical, electronic, and magnetic properties[11,12]. Most metal nitrides remain stable after decompression, making them particularly attractive targets for high-pressure synthesis. To date, high-pressure nitridation has primarily relied on two synthetic routes. The first employs direct reactions between elemental metals and molecular nitrogen in laser-heated diamond anvil cells[7,8]. Conventionally, this process is attributed to the thermal activation of the metal surface, which enables the direct reaction between metal species and molecular nitrogen. This reaction process is unselective. Along the same reaction pathway, metal azides may also form as competing products[13,14]. The second utilizes diffusion-controlled solid-state reactions between metal salts (MCl, M=metal element) and solid nitrogen sources ($NaN_3$, $Ca_3N_2$) in large-volume presses[15,16]. This process typically requires several hours to reach completion. This reaction characteristic is similar to those commonly observed in solid-state reactions under ambient conditions. In addition to conventional solid-state nitridation, metal nitrides can also be synthesized under ambient conditions through flux-mediated reaction pathways. Na flux ($NaNH_2$) has been employed as flux media to facilitate nitride synthesis, improving reaction kinetics and synthesis efficiency to some extent[17]. Nitride crystals are grown from Na flux. However, the solubility of nitrogen in Na flux salts remains inherently low under ambient-pressure condition. The reaction requires a continuous supply of nitrogen throughout the reaction process. Consequently, the reaction kinetics remain inherently limited.

In contrast to the two mechanisms described above, we previously identified a distinct route for nitride synthesis using metal oxides as precursors[18–20]. Although the reaction starts from solid-state powders, its behavior exhibits several striking features that differ fundamentally from those expected for conventional solid-state reactions[21], suggesting the involvement of an alternative reaction mechanism. Stable metal oxides such as $Fe_2O_3$, $Co_2O_3$ and NiO can be rapidly transformed into high-quality nitride crystals when boron nitride is employed as the nitrogen source. Compared with conventional nitridation routes, this reaction exhibits several anomalous features, including reaction times of only a few minutes, substantially reduced synthesis pressures and temperatures for noble-metal nitrides ($Re_xN$)[18], and efficient formation of multi-component transition-metal nitrides ($Fe_{3-x}Co_xN$, $Fe_{3-x}Ni_xN$)[22–24]. These features suggest the existence of a previously unrecognized high-pressure reaction mechanism of metal oxide-to-nitride conversion. However, owing to the lack of direct observation of the reaction process, the underlying pathway and kinetics remain unknown. Although in situ X-ray diffraction (XRD) has been used to investigate high-pressure chemical reactions [25], its capability is inherently limited when amorphous or liquid phases are involved, as these phases produce little or no long-range diffraction information. Moreover, in situ XRD primarily provides structural snapshots and cannot directly capture the dynamic evolution of high-pressure chemical reactions.

In this study, we use real-time synchrotron X-ray radiography to image this reaction within a large volume press, at the beamline BL12SW of

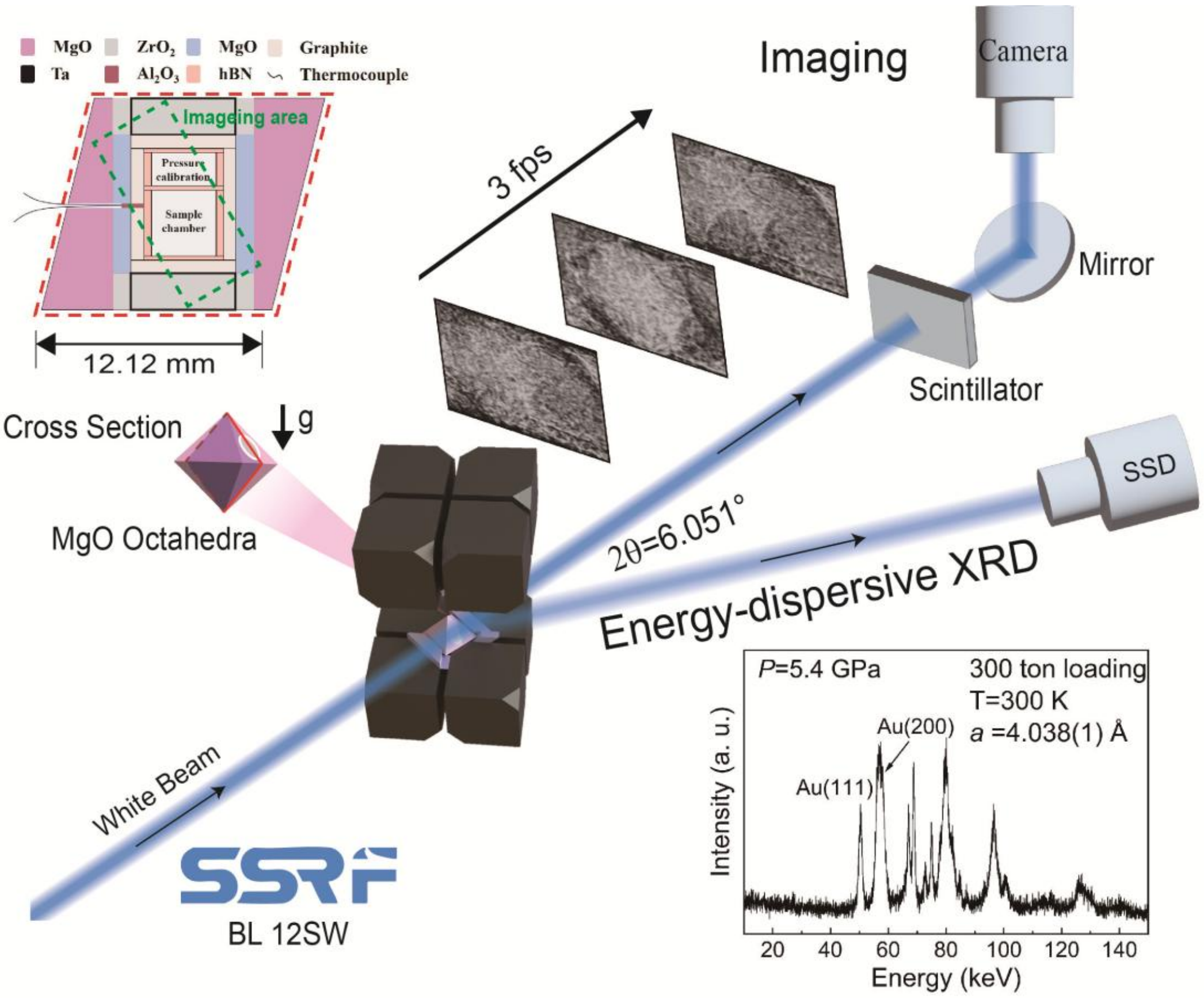

**Figure 1**. Schematic illustration of the real-time synchrotron X-ray radiography. The radiographic images are acquired using a camera. The size of the white beam is adjusted using a slit system composed of four tungsten carbide plates. During energy-dispersive X-ray diffraction measurements, the beam size is approximately 100 μm × 100 μm, and the diffraction data are collected using a solid-state detector (SSD). The lower-right panel shows the EDXRD pattern of the pressure standard Au under a load of 300 tons, with additional diffraction peaks arising from MgO and fluorescent peaks from Pb (Figure S1). The upper-left panel presents a schematic cross-sectional view of the sample assembly used in the pressure calibration experiment. The cross-sectional view is rotated by 45° relative to the octahedral assembly. The black downward arrow indicates the direction of gravity during imaging, referenced to the octahedral assembly.

Shanghai Synchrotron Radiation Facility (SSRF)[26,27]. Using $Fe_2O_3$ and BN as precursors, we capture the entire reaction process of a metal oxide into a metal nitride at 5.4 GPa and ~ 1700 K. It clearly revealed that this conversion involves a two-stage liquid-liquid separation process (LLPS) in borate melts. On the basis of these observations, we propose a nitrogen-driven mechanism for LLPS in borate melts.

**Real-time Synchrotron X-ray Radiography Experiments**

A schematic illustration of the real-time synchrotron X-ray radiography is shown in Figure 1. In this study, four real-time synchrotron X-ray experiments were conducted. The first experiment was used for pressure calibration (Exp. 0). The sample chamber was divided into two layers, with a mixture of nanoscale Au powder and MgO powder in the upper layer (Figure S2). The diffraction signal from the pressure standard region is collected using energy-dispersive X-ray diffraction (EDXRD). The pressure was calibrated based on the Au (111) diffraction peak of Au using its equation of state[28]. Under a loading force of 300 tons, the pressure in the chamber was determined to be 5.4 GPa. In the subsequent experiments (Exp. 1 and Exp. 2), the sample chamber was not divided into two layers. The sample chamber contained only a precursor powder mixture of $Fe_2O_3$ and BN. The same maximum loading force of 300 tons was applied in Exp.1 and Exp. 2. After reaching a loading force of 300 tons, *in-situ* observations of the sample chamber were conducted using X-ray radiography with a temporal resolution of 3 frames per second. The heating process was controlled by adjusting the input power manually. The temperature of the sample chamber is monitored in real time using a W–Re thermocouple (Figure S3). Full imaging videos are provided in the Supplementary Material. The acquired radiography images were subsequently processed using ImageJ software.

**Observation of Liquid-Liquid Phase Separation**

The imaging videos clearly reveal that, upon heating, fluid nitrogen and fluid Fe–N alloy sequentially separate from the system via LLPS. According to the Beer–Lambert law,

$$I = I_0 e^{-\rho k x} \tag{1}$$

$I$ and $I_0$ denote the transmitted and incident X-ray intensities, respectively, $\rho$ is the density of the penetrated material, $k$ is the absorption coefficient, and $x$ is the sample thickness. In radiography images, the pixel intensity (gray level) is closely related to the local absorption coefficient and effective density, such that variations in gray level can reflect changes in the local density. To quantitatively analyze the LLPS process, 12 representative pixels were selected at equal intervals within the imaging region, and their gray values were extracted and plotted as a function of temperature. Among them, points 1, 4 and 9 are located in the MgO pressure medium, where the gray values remain nearly constant throughout the heating process, showing no significant variation. The other points are located within the sample chamber. During the LLPS process, fluid nitrogen and the fluid Fe–N alloy appear in different regions. As a result, the temperature-dependent evolution of gray values varies significantly among different points. Overall, pronounced changes in gray values are observed at two characteristic temperature ranges, which correspond to the separation temperatures of fluid nitrogen and fluid Fe–N alloy from the system, respectively.

To further confirm whether the observed white spherical features in the radiography images correspond to high-pressure nitrogen fluid, we performed a quantitative analysis. Firstly, under high pressure, no voids are expected to exist within the sample chamber. Therefore, the formation of these voids can be excluded from being caused by volume changes associated with the solid-to-melt transition of the precursors. On the other hand, the existence of the fluid Fe–N alloy allows for a quantitative analysis of the system composition based on the relationships among volume, density, and molar ratio. The reaction process can be expressed as follows[20]:

$$3Fe_2O_3 + 6BN \rightarrow 2Fe_3N + 3B_2O_3 + 2N_2 \qquad (2)$$

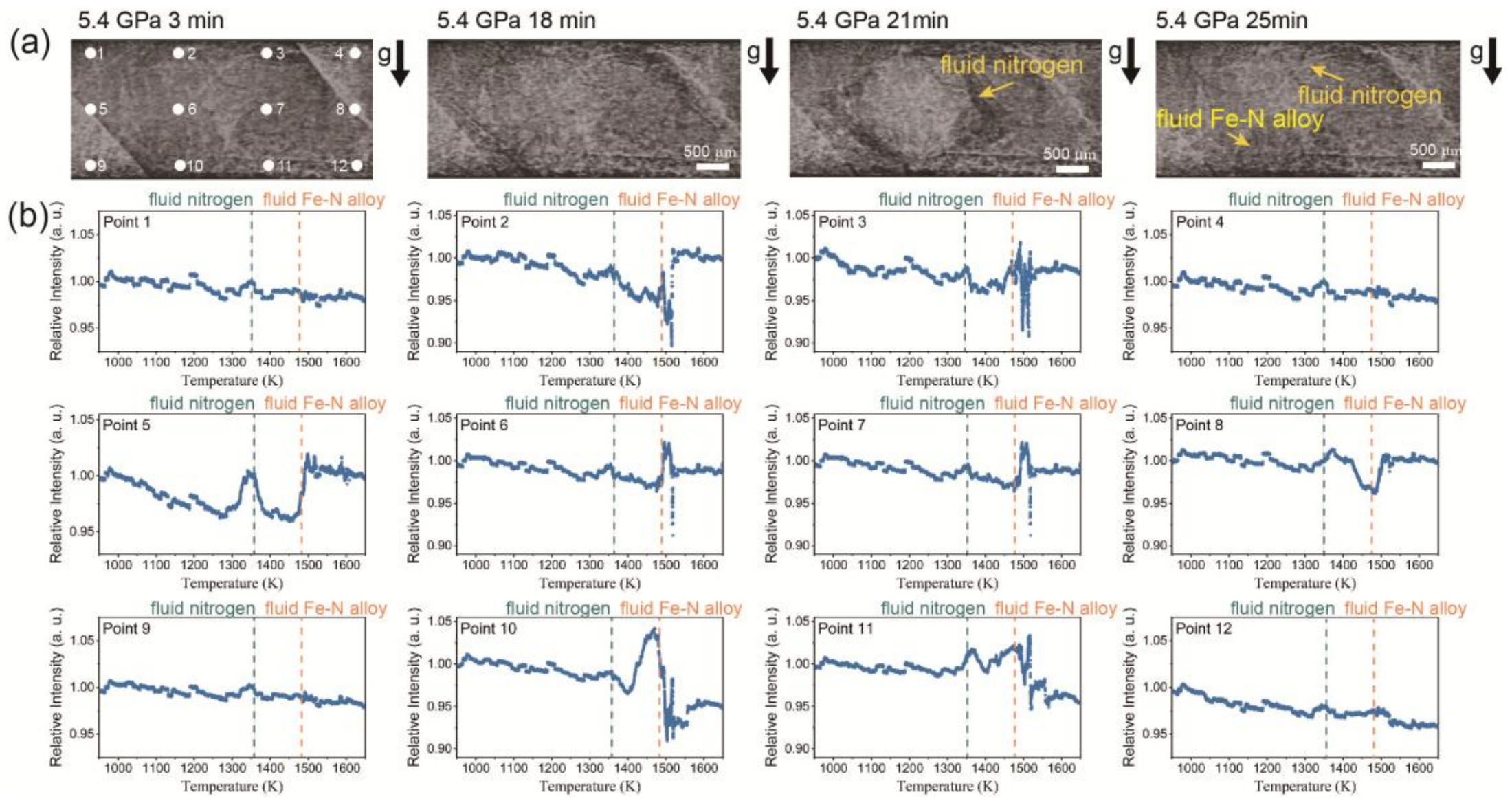

**Figure 2**. (a) Radiography images of the sample chamber at different stages in Exp. 1. The time stamps shown above correspond to those in the video provided in the Supplementary Information. The white circles indicate the regions selected for grayscale intensity analysis in panel (b), while the black arrows denote the direction of gravity. (b) Evolution of image intensity within the selected region as a function of temperature throughout the entire LLPS process. The green and orange dashed lines mark the onset temperatures of LLPS for fluid nitrogen and fluid Fe-N alloy, respectively.

The reaction equation has been discussed in detail in our previous work[18,19]. The characterizations of both the precursors and products are provided in the Supporting Information (Figure S4-S6). If the molar ratio between fluid nitrogen and fluid Fe–N alloy observed under high-pressure conditions approaches 1:1, this provides further evidence that the white spheres observed in the imaging correspond to fluid nitrogen.

The melting curves of the relevant matter are shown in Figure 3a. At approximately 5.4 GPa, the melting temperatures of Fe and $Fe_3N$ are ~1600 K and ~2000 K, respectively. In Exp.1, the maximum temperature reached was ~1700 K, which is significantly lower than the melting temperature of Fe. Therefore, the metallic fluid observed in the imaging videos, formed via LLPS, cannot be attributed to pure Fe melt, but instead exists as a fluid Fe–N alloy. Moreover, the temperature at which the fluid Fe–N alloy forms and persists as a spherical melt is consistent with the melting point of iron nitride previously determined from DAC experiments[29]. Figure 3b shows the density curve derived from the equation of state, which is fitted using a third-order Birch–Murnaghan (BM) equation [30]:

$$P = \frac{3}{2}K_0\left[\left(\frac{\rho}{\rho_0}\right)^{7/3} - \left(\frac{\rho}{\rho_0}\right)^{5/3}\right]\cdot\left\{1 + \frac{3}{4}(K' - 4)\left[\left(\frac{\rho}{\rho_0}\right)^{2/3} - 1\right]\right\} \qquad (3)$$

At 300 K and 5.4 GPa, the densities of $Fe_3N$ and solid nitrogen are 7.65(5) g/cm$^3$ and 1.61(8) g/cm$^3$, respectively. In the *in-situ* synchrotron imaging videos, a frame in which the nitrogen fluid forms a well-defined spherical shape was selected for diameter measurement (Figure S7). The sample chamber has a diameter of 3.5 mm under ambient conditions. Since the imaging field of view remains fixed, the diameters of the nitrogen and Fe–N alloy spheres were measured to be approximately 1.73(6) mm and 1.82(1) mm, respectively. Based on the spherical volume formula and the relations $m = \frac{4}{3}\pi R^3 \rho$, and $n = \frac{m}{Mr}$, wherer $n$ denotes the amount of substance, and $Mr$ represents relative molecular mass. The relevant parameters are summarized in Table S1. The resulting molar ratio is $n_{\mathrm{N_2}}: n_{\mathrm{Fe-N}} = 0.15(7): 0.13(3)$,which is in general agreement with the stoichiometry predicted by the reaction equation. The ~15% deviation can be attributed to the use of the density at room temperature of solid nitrogen and Fe3N in the calculation. The change in density upon melting is not identical for these two matters. On the other hand, some small spherical droplets of fluid nitrogen and Fe-N alloy did not coalesce into a larger sphere. This also contributes to the overall uncertainty in the analysis.

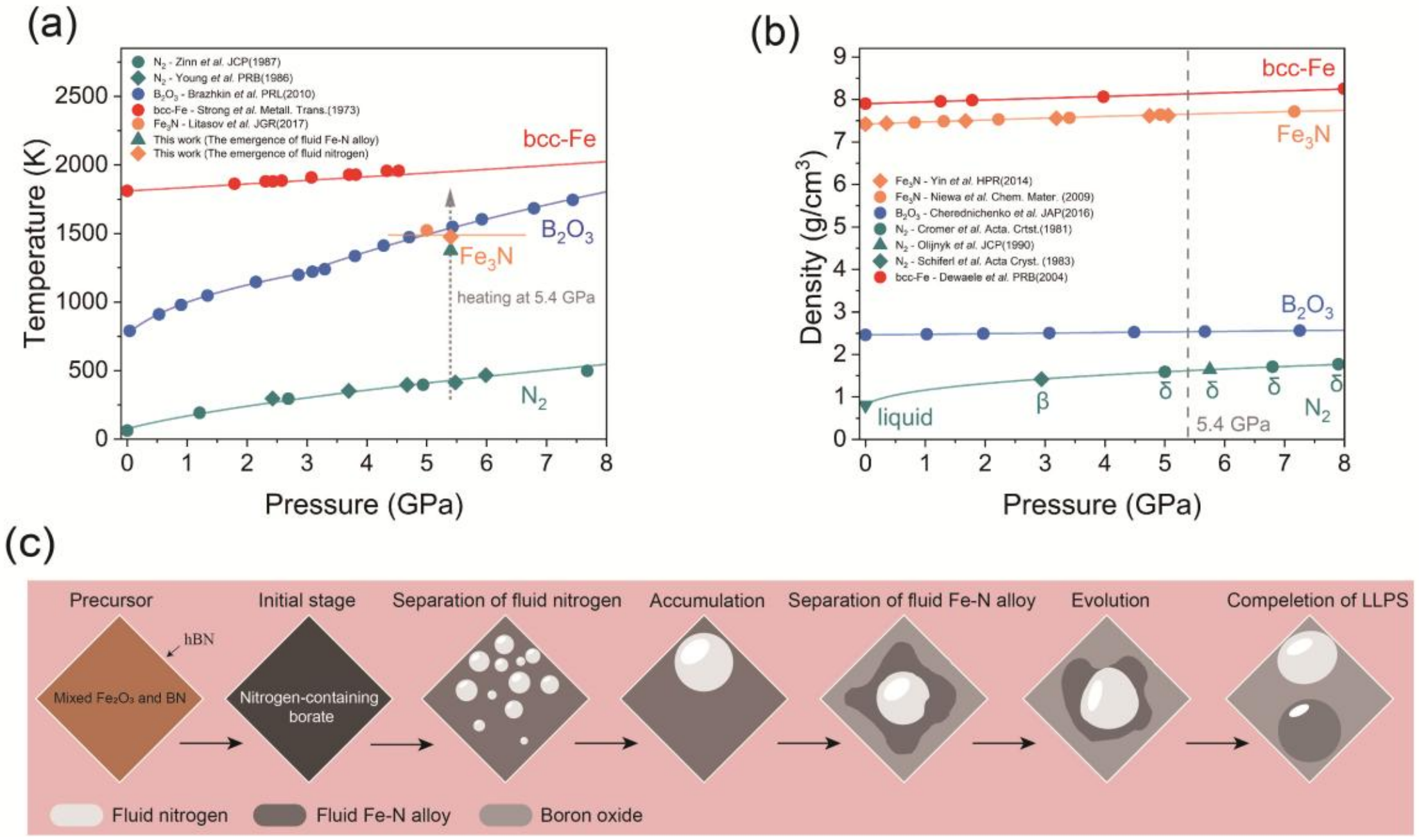

**Figure 3**. (a)Melting curves and (b) pressure-dependent densities of bcc-Fe[31,32], $B_2O_3$[33,34], $N_2$ [35–39], and $Fe_3N$[20,29,40]. Different solid symbols represent reference data from various literature sources. The melting curves are fitted using Simon–Glatzel equations, while the density–pressure relationships are described by a third-order Birch–Murnaghan equation of state. The gray dotted arrow in panel (a) indicates the P–T path followed in this study. (c) Schematic illustration of the LLPS mechanism constructed based on the radiographic observations. The pink background denotes the MgO pressure-transmitting medium used in the experiment. The white spheres represent fluid nitrogen, while the black spheres and the corresponding irregular shapes represent the fluid Fe–N alloy. The entire process can be divided into seven distinct stages.

Figure 3c schematically illustrates the kinetic evolution of the reaction. Notably, the phase behavior and dynamic features observed by synchrotron X-ray radiography are consistent with previous studies on the physical properties of molten $B_2O_3$ and high-pressure fluid nitrogen[34,41]. As illustrated in Figure 3(c), the mixed $Fe_2O_3$ and BN precursors react to form a borate. At approximately 1400 K, small amounts of fluid nitrogen in form of small bubbles appear at multiple locations within the sample chamber. These droplets rise and coalesce at the top of the chamber, forming a larger fluid nitrogen sphere. With increasing temperature, the fluid Fe–N alloy begins to throughout the sample chamber, the fluid Fe–N alloy gradually migrates and accumulates toward separate via LLPS. Initially distributed the center. At this stage, the contraction associated with Fe–N aggregation drives the fluid nitrogen toward the center of the chamber. The presence of fluid nitrogen interferes with the coalescence of the fluid Fe–N alloy, leading to the formation of a core–shell-like configuration. During this stage, the fluid Fe–N alloy (shell) exerts compressive forces on the fluid nitrogen (core). Under this compression, the fluid nitrogen exhibits noticeable elasticity, deforming while maintaining its integrity without breaking into smaller droplets, suggesting a certain degree of liquid elasticity. This behavior is associated with the orientational correlations in supercritical fluid nitrogen reported by Temleitner et al.[41]. Under these conditions, the fluid molecules exhibit enhanced intermolecular interactions. Eventually, the two phases become fully separated. In this stage, fluid nitrogen exhibits a relatively fast upward migration within the sample chamber, indicating a relatively low interfacial resistance. Notably, $B_2O_3$ is an archetypal super-viscous glass-forming liquid. Previous studies have shown that its viscosity decreases by approximately four orders of magnitude at ~5.5 GPa relative to ambient pressure, approaching that of low-viscosity liquids as olive oil[34].

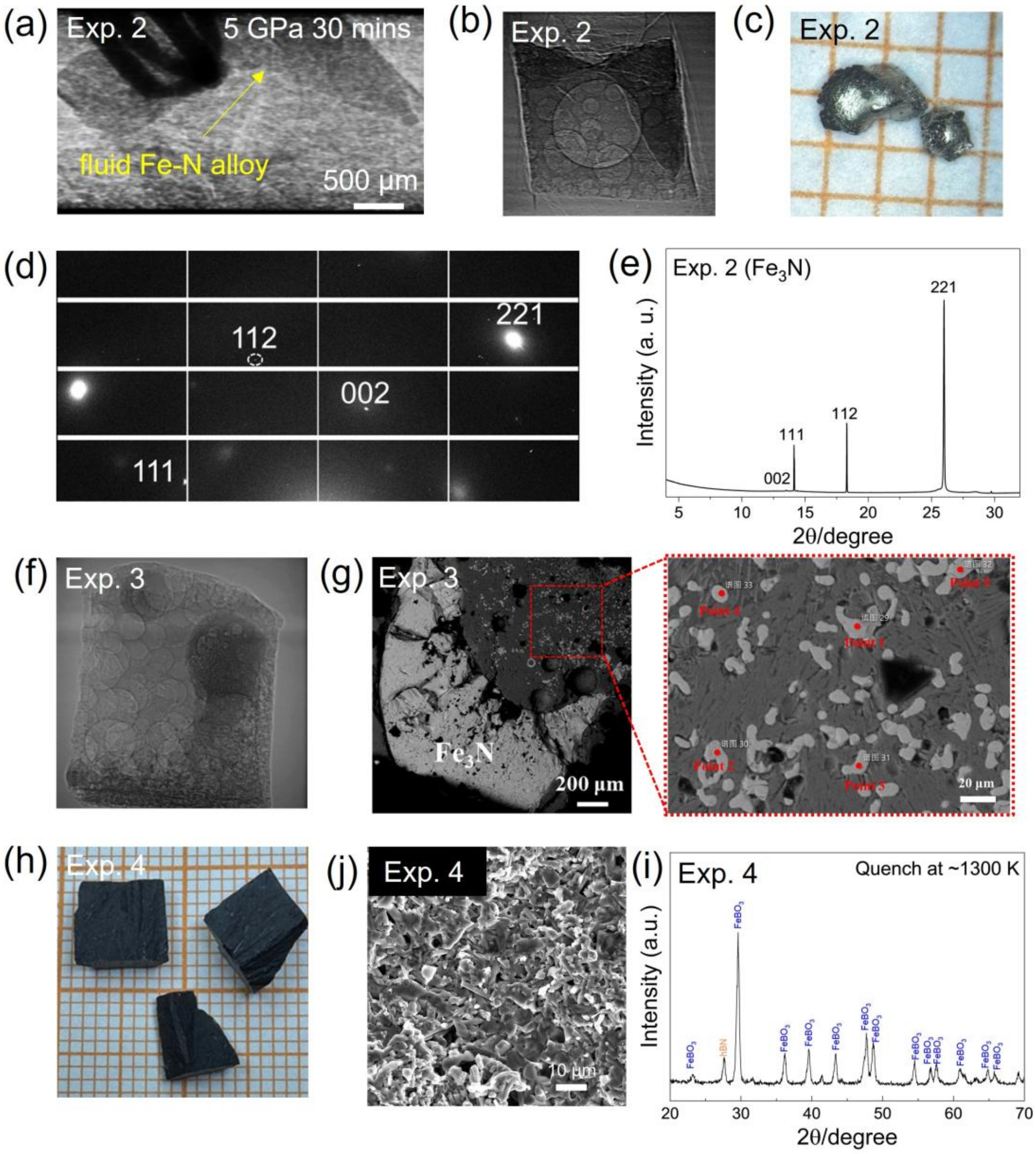

**Figure 4.** In Exp. 2 and Exp. 3, the samples were quenched immediately after the fluid Fe–N alloy formation was observed.(a) Synchrotron radiography image prior to quenching at 5.4 GPa.(b) Radiography image of the sample recovered to ambient conditions and after removal from the assembly.(c) Optical photograph of the $Fe_3N$ product.(d) Synchrotron angle-dispersive X-ray diffraction pattern and (e) corresponding diffraction image of the $Fe_3N$ product. (f) Radiography image of recovered sample from Exp. 3. (g) The EDS pattern of cross section after polishing. In the off-line experiment of Exp. 3, the sample was quenched at approximately 1300 K, prior to the onset of liquid–liquid phase separation (LLPS). (h) Optical image of the quenched sample. (i) Scanning electron microscopy image of the recovered sample, in which no pore structures associated with nitrogen are observed. (j) X-ray diffraction pattern of the quenched sample.

**Identification of Intermediates in Metal Oxide-to-Nitride Conversion**

Conventional solid-state nitridation reactions are generally limited by sluggish solid-state diffusion and often require several hours or even longer to reach completion. In contrast, our results reveal that the formation of the fluid Fe–N alloy in this pathway is completed within only a few minutes, representing an enhancement in reaction rate of more than one order of magnitude compared with conventional solid-state nitridation. Such a pronounced kinetic disparity indicates that the reaction does not follow a classical diffusion-controlled solid-state mechanism, but instead proceeds through a fundamentally distinct high-pressure reaction pathway. To elucidate the underlying microscopic mechanism, quenched samples recovered from different reaction stages were systematically characterized in combination with real-time synchrotron X-ray radiography.

To determine the phase evolution at different reaction stages, in Exp. 2 the sample was quenched immediately after the formation of the fluid Fe–N alloy formation was observed. The video is also provided in the Supporting Information. Figure 4c shows synchrotron radiography images of the recovered sample after removal of the encapsulation material at ambient conditions. Owing to the elimination of absorption from the surrounding pressure medium, the image exhibits significantly enhanced contrast. Notably, $Fe_3N$ retains an irregular, fluid-like morphology that developed under high pressure and high temperature (Figure 4b). In addition, voids are observed within the sample chamber, whose spatial distribution closely matches the regions of fluid nitrogen identified under high pressure. These cavities are therefore interpreted as being generated by nitrogen exsolution during decompression, providing further evidence for the existence of a fluid nitrogen phase at high pressure and temperature. The quenched products were further characterized by synchrotron X-ray diffraction (Figure 4d and 4e). The diffraction peaks can be indexed to $Fe_3N$. Owing to the micron-scale size of the synchrotron beam, pronounced preferred orientation effects are observed in the diffraction patterns. In Exp. 3, the sample was also quenched immediately after the formation of the liquid Fe–N alloy. In contrast to Exp. 2, the by-products were retained, and cross-sections were exposed by epoxy mounting and polishing. The transmission images of the recovered sample at ambient conditions are shown in Figure 4f. A large number of diffusely distributed droplet-like particles are observed in the upper-right region of the sample. Combined with EDS analysis (Table S2), these particles are identified as the $Fe_3N$ phase. Notably, these $Fe_3N$ particles remain isolated and have not coalesced into larger droplet structures. Together, the results from Exp. 2 and Exp. 3 indicate that the Fe–N fluid is directly exsolved from the borate melt via a liquid–liquid phase separation process, rather than forming through initial precipitation of metallic Fe followed by subsequent nitridation via nitrogen diffusion.

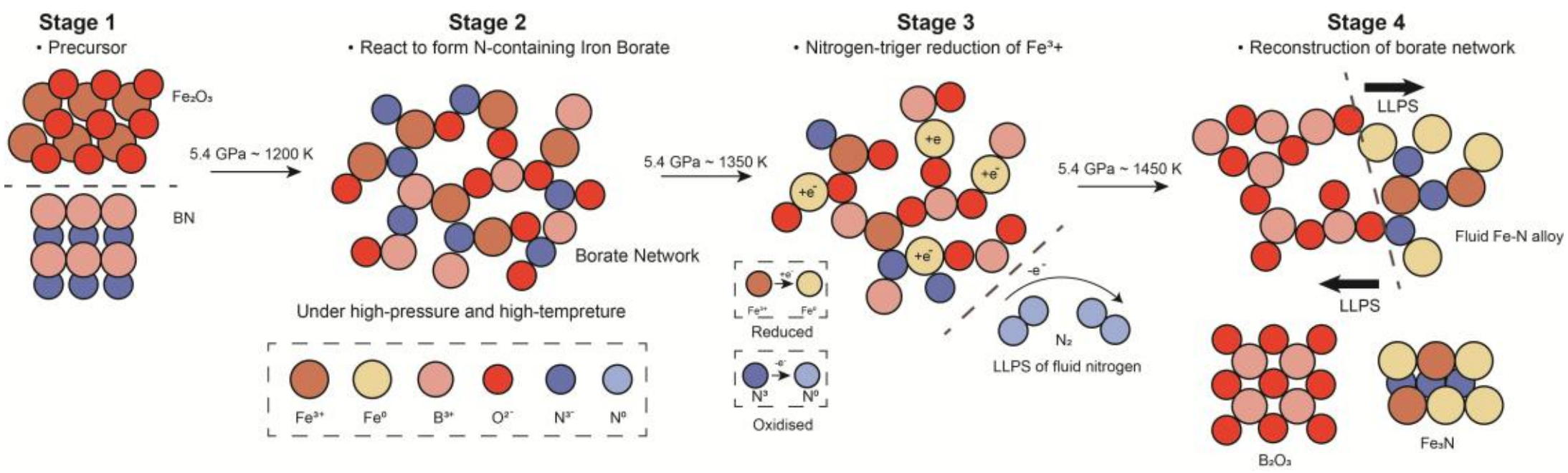

**Figure 5.** Schematic illustration of the nitrogen-driven borate melts reconstruction and liquid–liquid phase separation (LLPS) mechanism. The first stage shows a physical mixture of $Fe_2O_3$ and BN precursors. In the second stage, $Fe_2O_3$ reacts with BN to form a nitrogen-containing borate melt. In the third stage, nitrogen is oxidized to form a fluid molecular nitrogen phase. In the fourth stage, changes in the coordination environment of Fe induce reconstruction of the borate network, leading to network destabilization and subsequent liquid–liquid phase separation (LLPS), ultimately resulting in the separation of fluid Fe–N alloy.

To further capture intermediates at the early stage of the reaction, in Exp. 4 (off-line experiment), the experiment was quenched prior to the onset of the first liquid–liquid phase separation event. SEM reveals that no pore structures corresponding to fluid nitrogen are present within the sample (Figure 4i), indicating that fluid nitrogen has not yet formed at this stage. X-ray diffraction analysis shows that $FeBO_3$ has already formed (Figure 4j), while the diffraction peaks of the initial $Fe_2O_3$ precursor disappear (Figure S1). In addition, weak diffraction signals corresponding to unreacted BN are still detectable. Since the subsequent liquid–liquid phase separation requires the involvement of nitrogen, and no independent nitrogen phase is detected at this stage, it can be inferred that nitrogen has already been incorporated into the borate framework under high-pressure conditions, leading to the formation of a nitrogen-containing borate melt. However, this metastable nitrogen-bearing configuration is not preserved upon decompression to ambient conditions. Instead, $FeBO_3$ is retained as the observable quenched product.

**Mechanism of Borate Melt-Mediated Metal Oxide-to-Nitride Conversion**

The experimental observations presented above establish borate species as key intermediates in the oxide-to-nitride conversion pathway. The reaction mechanism shown in Figure 5 is proposed. Initially, the metal oxides react with BN to form a nitrogen-containing borate melt. Previous studies have shown that borates under high-pressure and high-temperature conditions typically exist as glassy[34,42]. As a consequence, even *in-situ* energy-dispersive X-ray diffraction (EDXRD) cannot resolve their local structural motifs or unambiguously determine the speciation of nitrogen. At this stage, no metal nitrides are formed, and metal cations remain structurally embedded within the borate framework. Upon further heating to approximately 1300 K, nitrogen begins to participate in the reduction of metal cations, accompanied by the gradual formation of fluid nitrogen. Concurrently, a fraction of the metal cations is reduced to lower oxidation states or to the metallic state. Because reduced and oxidized metal species possess distinctly different coordination environments, the original structural balance of the borate network is disrupted, thereby triggering network reconstruction. During this reconstruction process, nitrogen exhibits a stronger affinity toward the metallic phase compared with oxygen, leading to progressive nitrogen enrichment in the metal-rich domains. This ultimately drives the exsolution of a fluid Fe–N alloy from the borate melt, forming an independent liquid phase.

Using the same synthetic strategy, we previously synthesized $Fe_{3-x}Co_xN$ and $Fe_{3-x}Ni_xN$ from mixed $Fe_2O_3$/CoO and $Fe_2O_3$/NiO oxide precursors with BN, respectively[22,24]. Despite the differences in composition of precursor, all recovered products exhibit similar structural features, including nitrogen-derived voids and spherical metal nitride. The consistent appearance of these characteristic features suggests that these systems likely share a common reaction pathway, supporting the broader applicability of the borate-mediated oxide-to-nitride conversion mechanism proposed in this work. The mechanistic picture established in this work allows our previously reported oxide-to-nitride synthesis systems to be reinterpreted from a unified perspective. Firstly, the formation of metal nitrides does not proceed through a conventional diffusion-controlled solid-state pathway. Instead, it is mediated by structural reconstruction within the melt and a subsequent liquid–liquid phase separation process, which accounts for the significantly enhanced reaction kinetics compared with traditional solid-state routes. Second, this mechanism also rationalizes the unique capability of the present approach to synthesize multicomponent metal nitrides. The nitrogen-containing borate melt serves as a critical intermediate phase throughout the reaction.

$$\{M \cdot [BO] + N \cdot [BO]\}_{melt} \xrightarrow{HPHT} MN + B_2O_3 \quad (5)$$

Under high-pressure and high-temperature conditions, the B–O network can adopt a wide range of local structural motifs[43,44], endowing the borate melt with exceptional solvation ability that enables the simultaneous incorporation and homogeneous mixing of multiple metal cations. Consequently, during liquid–liquid phase separation, different metal species can be collectively partitioned into the nitrogen-rich metallic phase, thereby providing a favorable pathway for the formation of multicomponent transition-metal nitrides.

$$\{Fe \cdot [BO] + Co \cdot [BO] + N \cdot [BO]\}_{melt} \xrightarrow{HPHT} Fe_{3-x}Co_xN + B_2O_3 \quad (6)$$

$$\{Fe \cdot [BO] + Ni \cdot [BO] + N \cdot [BO]\}_{melt} \xrightarrow{HPHT} Fe_{3-x}Ni_xN + B_2O_3 \quad (7)$$

Collectively, these results indicate that nitrogen-containing borate phases serve as key intermediates in the oxide-to-nitride conversion pathway, providing a common mechanistic link between the synthesis of multicomponent transition-metal nitrides.

In a conclusion, real-time synchrotron X-ray radiography was employed to directly visualize a high-pressure chemical reaction and elucidate the pathway of metal oxide-to-nitride conversion. The results demonstrate that this transformation does not proceed through a conventional diffusion-controlled solid-state mechanism but instead involves a two-stage LLPS process in borate melts. We propose a nitrogen-driven mechanism for LLPS in borate melts. Our work provides the direct mechanistic insight into metal oxide-to-nitride conversion under high pressure and establishes a previously unrecognized reaction pathway for nitride formation. More importantly, it reveals how high pressure can create fundamentally new reaction pathways that are inaccessible under ambient conditions, opening new opportunities for mechanistic studies in high-pressure chemistry.

## Supporting Information

Complete *in-situ* synchrotron X-ray radiography videos of Exp. 1 and Exp. 2; pressure and temperature calibration details for real-time synchrotron X-ray radiography experiments; characterization of the precursors, products, and byproducts, including

X-ray diffraction (XRD) and energy-dispersive X-ray spectroscopy (EDS) results.

## Acknowledgments

This work was financially supported by the National Natural Science Foundation of China (Grant No. 12374013, Grant No. U2030107) and the Fundamental Research Funds for the Central University (Grant No. 2020SCUNL107). The real-time synchrotron X-ray radiography experiments were conducted at BL12SW of SSRF (Proposal No. 2024-SSRF-PT-505499, 2024-SSRF-JJ-505702).

# Supplementary Information for "Observation of Iron Oxide-to-Nitride Conversion via Liquid–Liquid Phase Separation in High-pressure Borate Melt"

Yu Tao[a], Depu Liu[a], Chunyin Zhou[b], Xu Jia[a], Jingyi Liu[a], Xue Chang[a], Yangbin Wang[a], Yipeng Wang[b], Duanwei He[1], Li Lei[a*]

*a. Institute of Atomic and Molecular Physics, Sichuan University, Chengdu 610065, China*

*b. Shanghai Advanced Research Institute, Chinese Academy of Sciences, Shanghai 201204, China*

*Electronic mail: lei@scu.edu.cn

**Sample Preparation and Characterization**

Reaction precursors consisted of $Fe_2O_3$ (Alfa Aesar, 99.99%) and hBN (Alfa Aesar, 99.99%). The powder was ground and mixed at a mass ratio of 1:10 for 30 min. The product and by-product were characterized by scanning electron microscopy (SEM, JSM-IT500HR, JEOL), energy-dispersive X-ray spectroscopy (EDX, INCA E250, Oxford Instruments), powder X-ray diffraction (XRD, DX-2700). Synchrotron angle-dispersive X-ray diffraction (ADXRD) was obtained from the Shanghai Synchrotron Radiation Facility (SSRF) BL15U1 beamline station.

**Radiographic imaging videoes**

The videoes show the complete metal oxide-to-nitride conversion, as observed in-situ using synchrotron X-ray radiography during heating to 1700 K at 5.4 GPa. The central region of the video corresponds to the sample chamber, surrounded by MgO as the pressure-transmitting medium. The white arrows in the video indicate the locations of fluid nitrogen and the fluid Fe-N alloy. A schematic illustration of the imaging area is provided in Figure 1 of the main text. The original video has a duration of approximately 30 min. To more clearly visualize the high-pressure chemical process, the video has been accelerated by a factor of 60. The timestamp displayed in the lower-right corner corresponds to the real (unaccelerated) time.

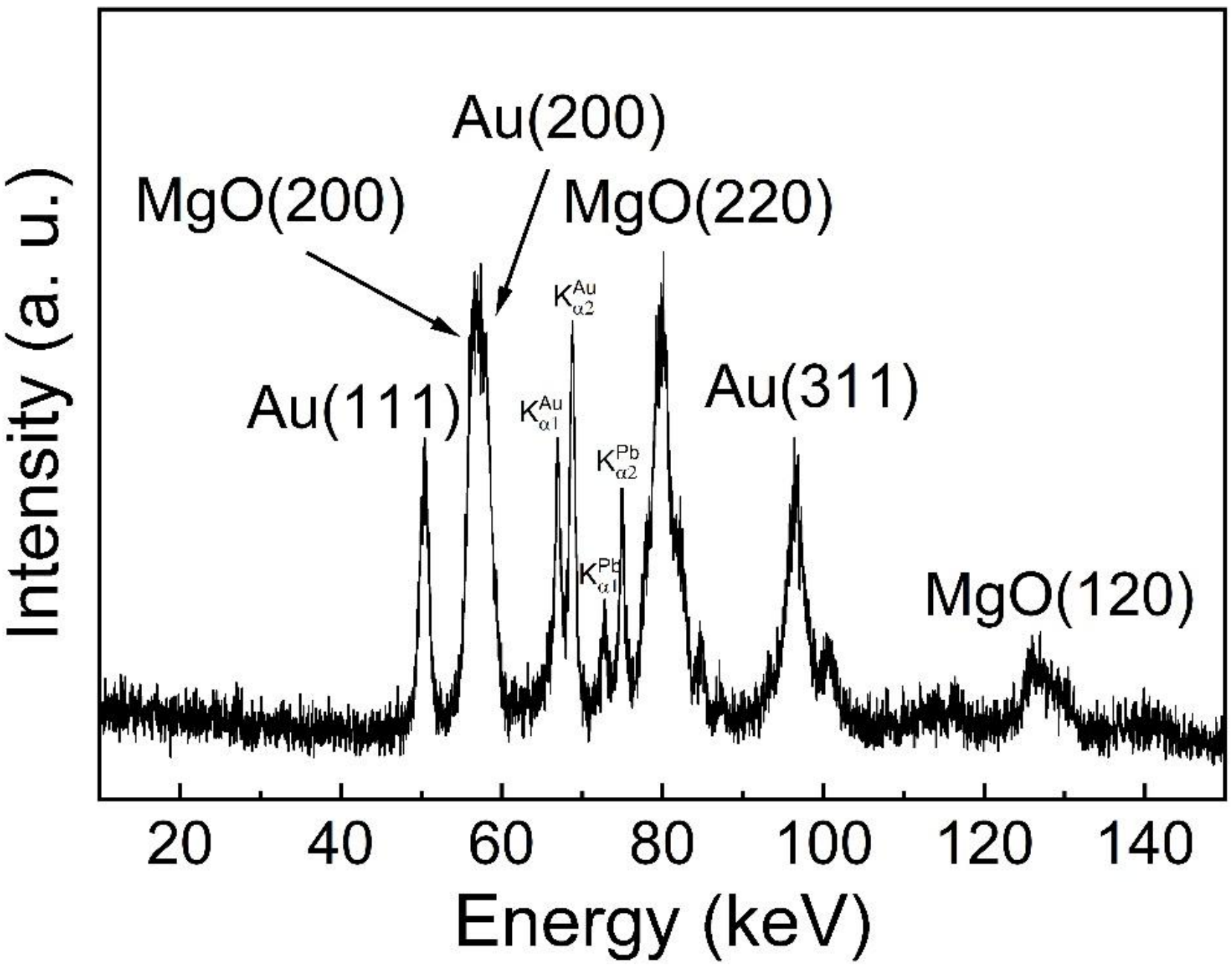

**Figure S1**. Peak identification in the ED-XRD pattern.

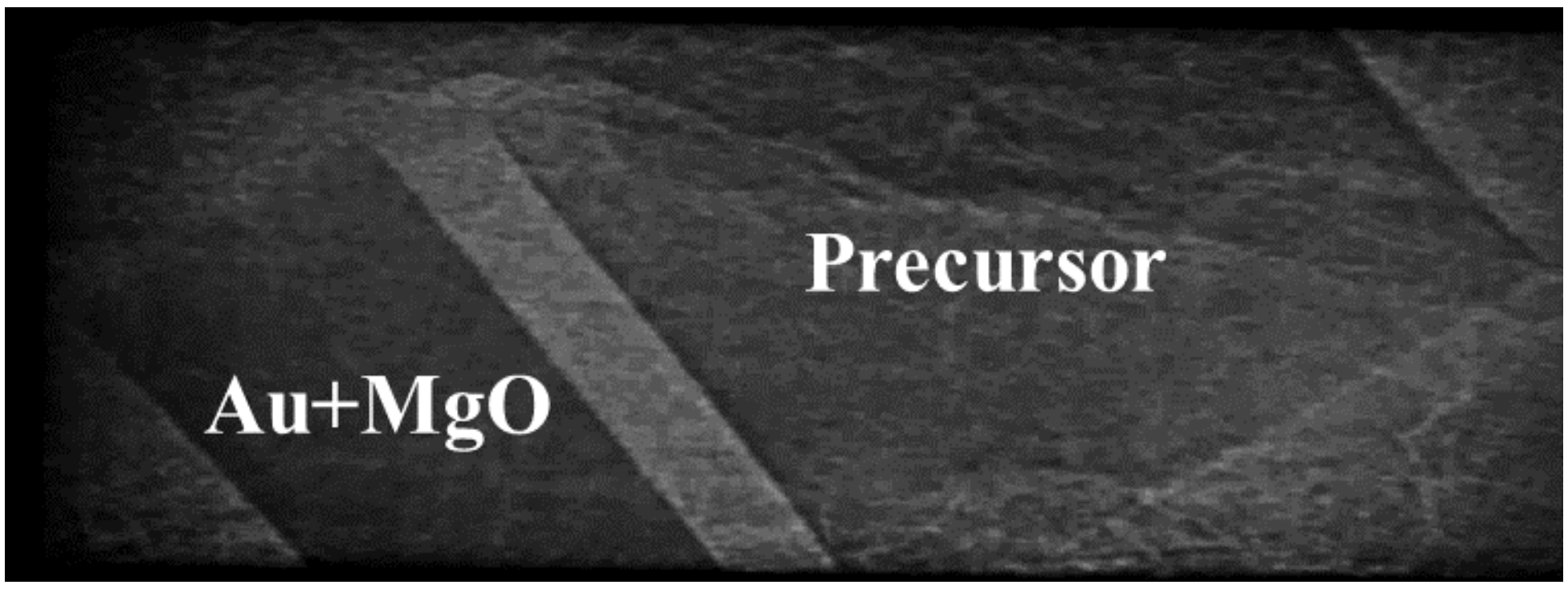

**Figure S2**. Radiographic image for chamber pressure calibration under a 300-ton press load in Exp. 0. A mixture of micron-sized Au and MgO powders was used as the pressure calibrant.

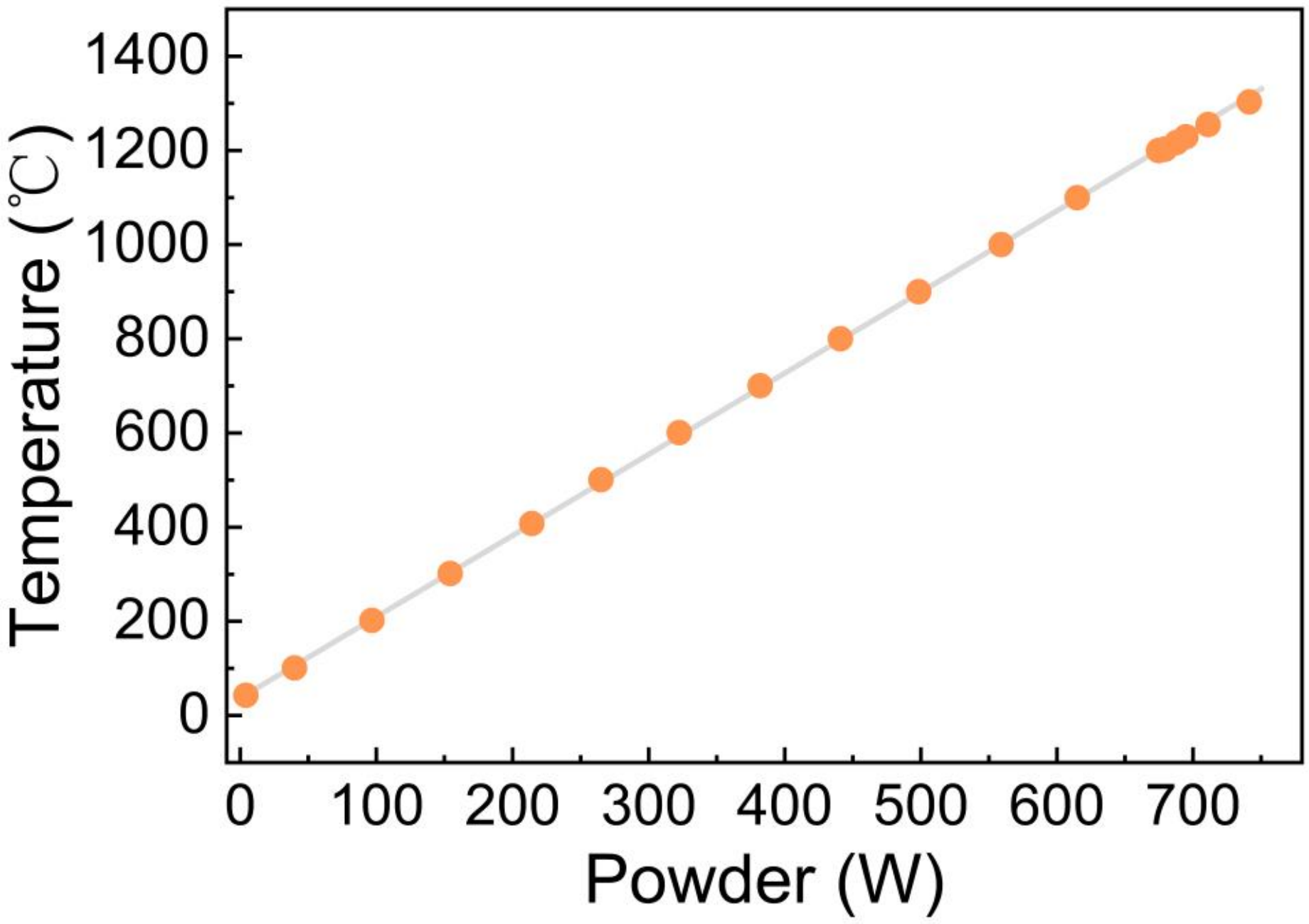

**Figure S3**. Temperature–power relationship of the sample chamber monitored *in-situ* using a W–Re thermocouple.

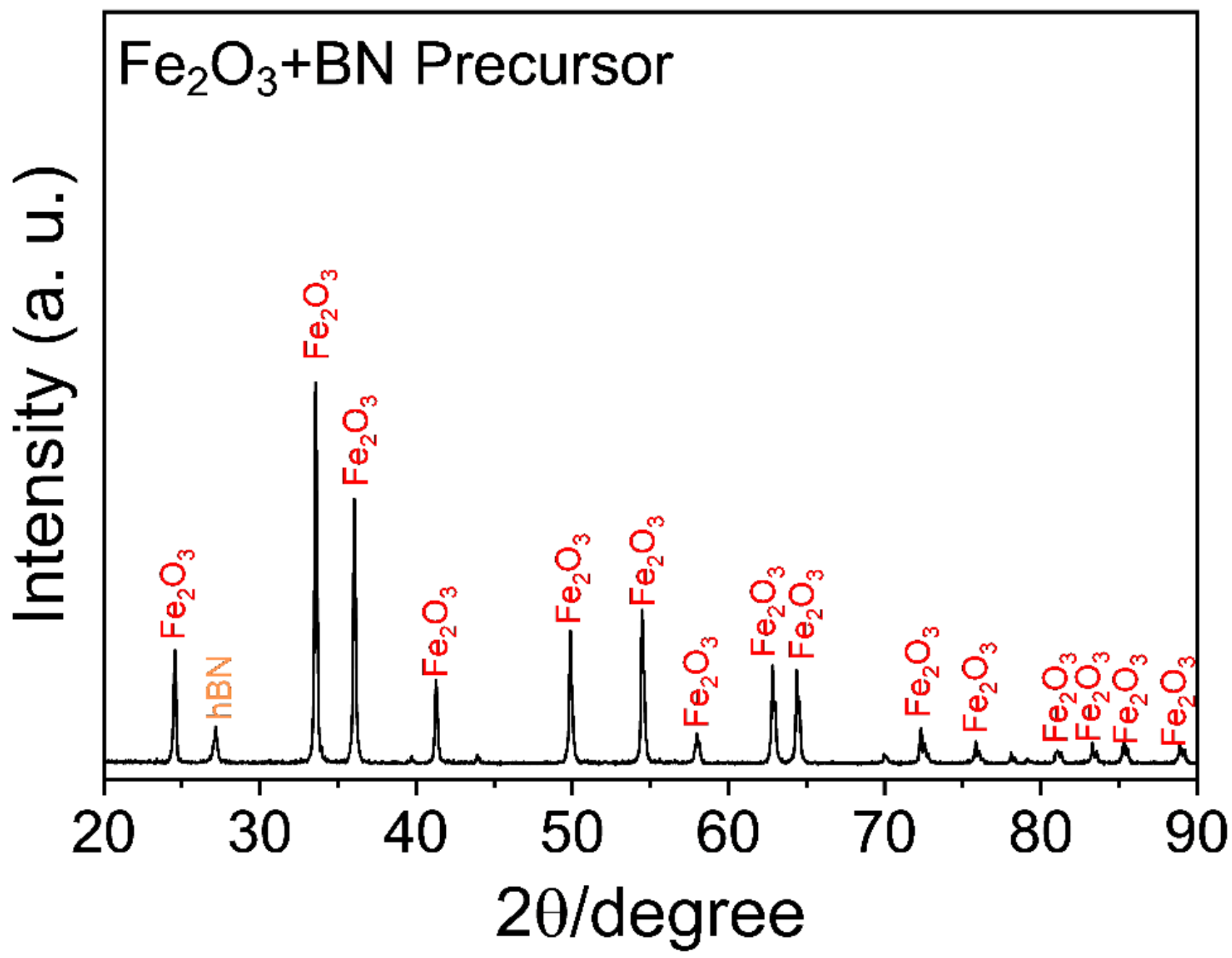

**Figure S4**. X-ray diffraction pattern of the mixed $Fe_2O_3$ and BN precursors prior to the high-pressure experiment.

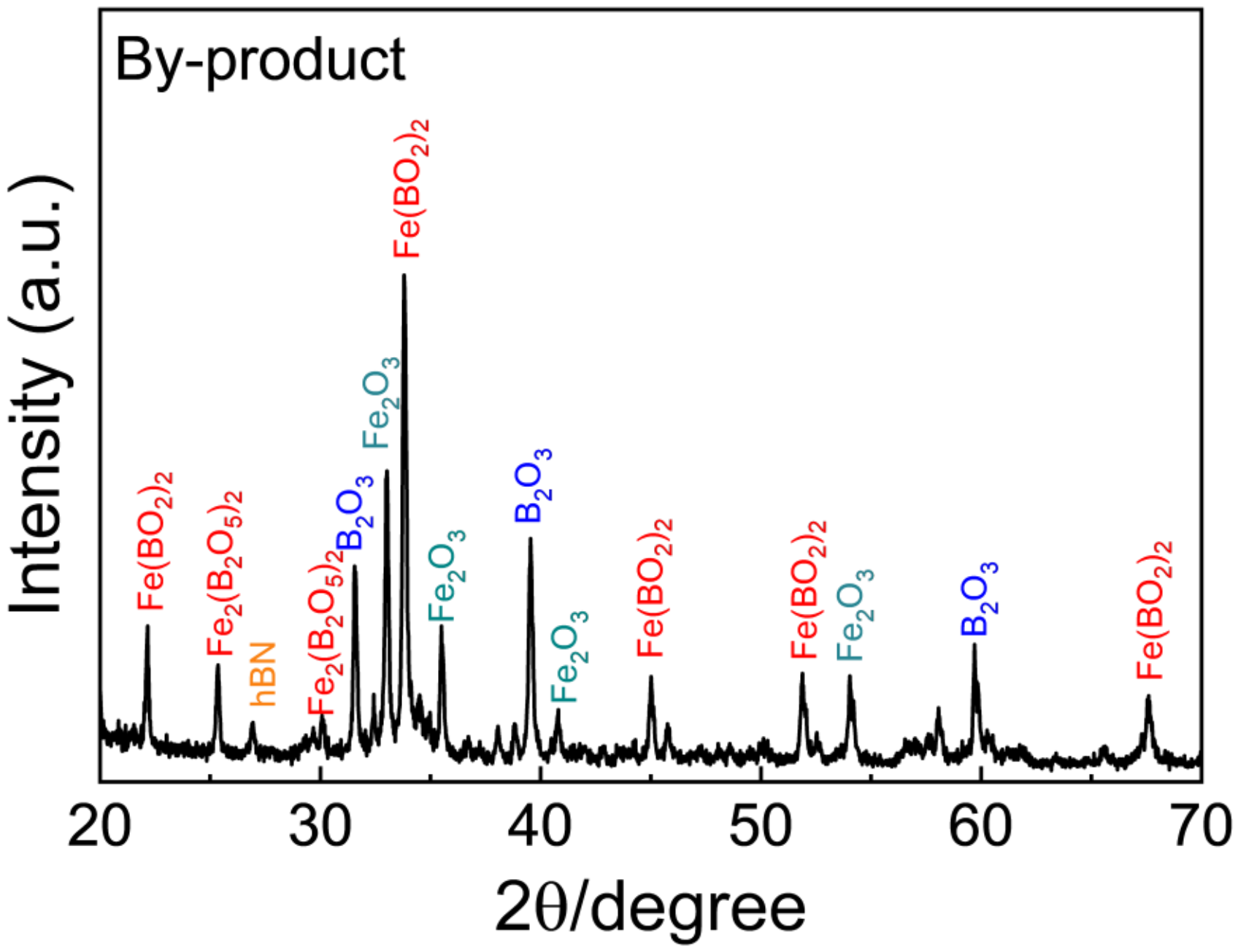

**Figure S5**. Powder X-ray diffraction pattern of the byproducts recovered after the reaction. The byproducts were characterized after grinding and drying at 60 °C for 12h. In addition to $B_2O_3$, the major byproduct in the reaction equation, $Fe(BO_2)_2$ was also identified. Its presence is attributed to the secondary reaction between residual $Fe_2O_3$ and the produced $B_2O_3$. Excess $Fe_2O_3$ is typically employed in the synthesis

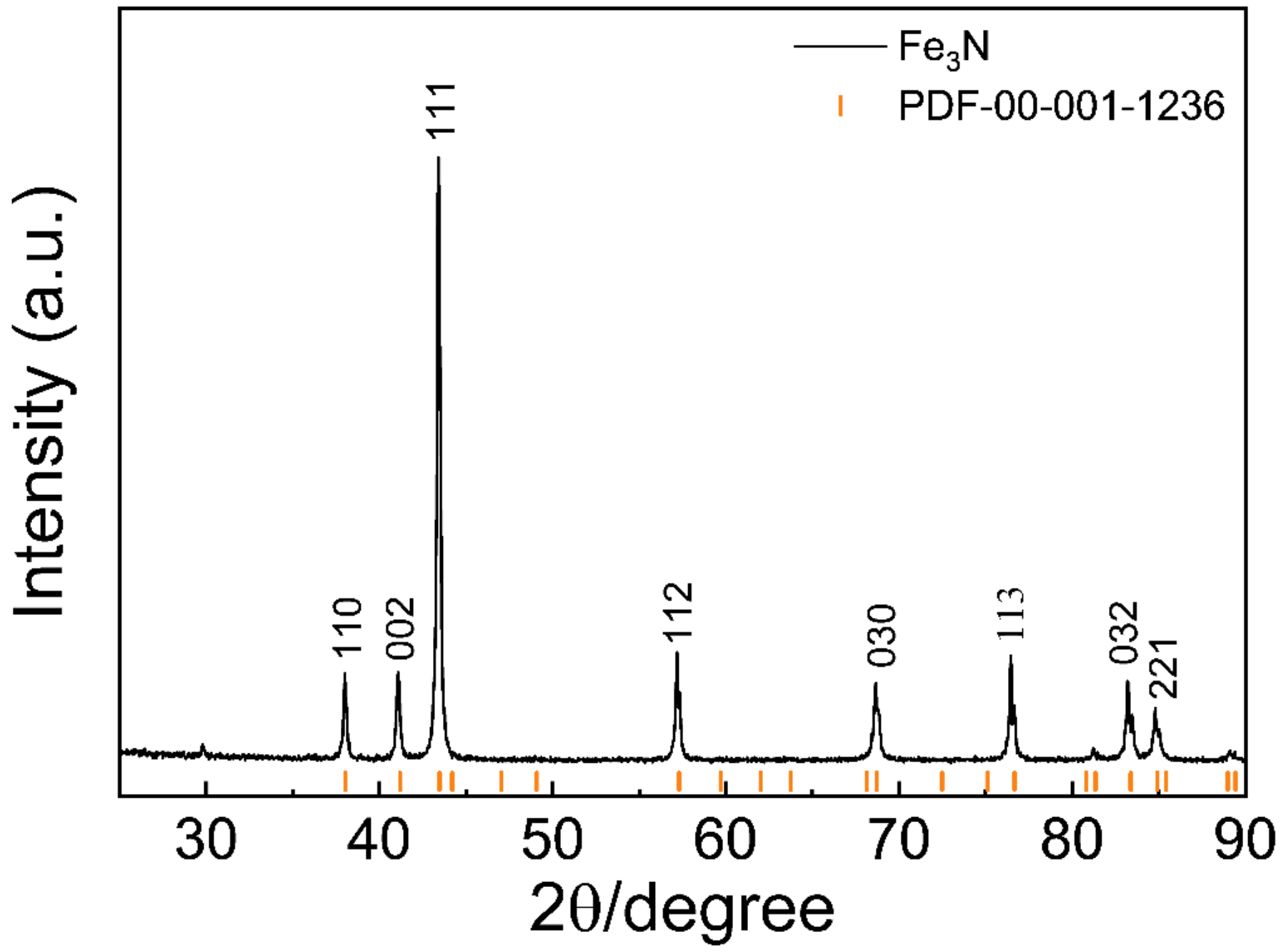

**Figure S6**. Powder X-ray diffraction pattern of the $Fe_3N$ product.

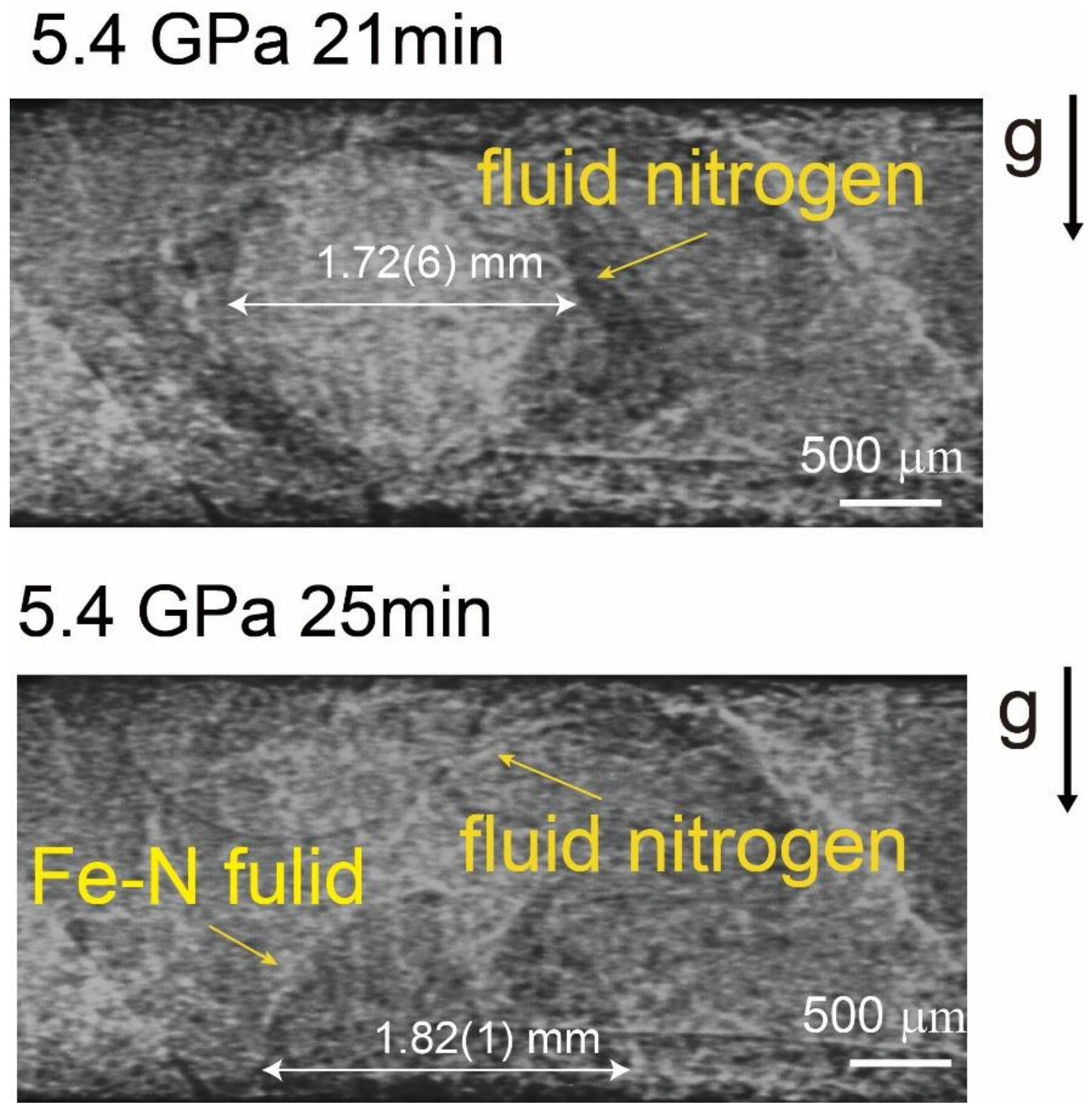

**Figure S7**. Diameter measurement of fluid Fe–N alloy and fluid nitrogen in Exp. 1.

**Table S1**. Masses and amount of substance of high-pressure fluid nitrogen and the fluid Fe–N alloy measured and calculated based on radiographic images.

| | *d* (*mm*)[a] | *V* (*mm*$^3$) | *ρ* (*g/cm*$^3$)[b] | *m* (*mg*) | *Mr* | *n* (*mmol*) |
|---|---|---|---|---|---|---|
| Fluid nitorgen | 1.73(6) | 2.71(1) | 1.61(8) | 4.38(6) | 28.02 | 0.15(7) |
| Fluid Fe-N alloy | 1.82(1) | 3.15(7) | 7.65(5) | 24.15(1) | 181.56 | 0.13(3) |

***a*** *The measured sphere diameters at 5.4 GPa and ~1700 K.*

***b*** *The density used in the calculation is obtained from a third-order Birch–Murnaghan equation of state fitted at 5.4 GPa and 300 K.*

**Table S2**. The EDS of product in Figure 4g. The sample surface was Pt-coated prior to SEM/EDS measurements.

| Lable | atom% | | |
|---|---|---|---|
| | Fe | N | Total |
| #1 | 76.5 | 23.5 | 100 |
| #2 | 75.6 | 24.4 | 100 |
| #3 | 77.7 | 22.3 | 100 |
| #4 | 77.9 | 22.1 | 100 |
| #5 | 81.0 | 19.0 | 100 |
| **Average** | 77.4(4)% | 22.2(6)% | |